\documentclass[%
 reprint,
 amsmath,amssymb,
 aps,
]{revtex4-2}

\usepackage{graphicx}% Include figure files
\usepackage{dcolumn}% Align table columns on decimal point
\usepackage{bm}% bold math

\usepackage{braket}
\usepackage[colorlinks,linkcolor=blue,anchorcolor=blue,citecolor=blue,urlcolor=blue]{hyperref}
\begin{document}

\preprint{APS/123-QED}

\title{Quantum Zeno Effect in the Spatial Evolution of a Single Atom}% Force line breaks with \\
\author{Zheng-Yuan Zhang$^{1,2}$}
\thanks{These authors contributed equally to this work.}
\author{Han-Chao Chen$^{1,2}$}
\thanks{These authors contributed equally to this work.}
\author{Xin Liu$^{1,2}$}
\author{Li-Hua Zhang$^{1,2}$}
\author{Bang Liu$^{1,2}$}
\author{Shi-Yao Shao$^{1,2}$}
\author{Jun Zhang$^{1,2}$}
\author{Qi-Feng Wang$^{1,2}$}
\author{Qing Li$^{1,2}$}
\author{Yu Ma$^{1,2}$}
\author{Tian-Yu Han$^{1,2}$}
\author{Ya-Jun Wang$^{1,2}$}
\author{Dong-Yang Zhu$^{1,2}$}
\author{Jia-Dou Nan$^{1,2}$}
\author{Yi-Ming Yin$^{1,2}$}
\author{Qiao-Qiao Fang$^{1,2}$}
\author{Dong-Sheng Ding$^{1,2}$}
\email{dds@ustc.edu.cn}
\author{Bao-Sen Shi$^{1,2}$}

\affiliation{$^1$Laboratory of Quantum Information, University of Science and Technology of China, Hefei 230026, China.}
\affiliation{$^2$Synergetic Innovation Center of Quantum Information and Quantum Physics, University of Science and Technology of China; Hefei, Anhui 230026, China.}

\date{\today}% It is always \today, today,
\maketitle

%\tableofcontents

\textbf{The quantum Zeno effect (QZE) reveals that frequent measurements can suppress quantum evolution, but the detailed dynamics of the system under finite-duration measurements in experiments remain insufficiently explored. Here, we employ an optical dipole trap as a projective measurement to study the motion of a single cold atom in free space. By monitoring atomic loss, we directly observe the QZE in single-atom motion in free space and find that the effect of dipole measurements on the atom comprises a short-time collapse followed by subsequent periodic unitary evolution, thereby providing an intuitive physical picture of measurement backaction across different timescales. We further investigate the effects of measurement frequency, strength, and spatial position, demonstrating that measurements not only suppress the spreading of quantum states but also enable deterministic preparation of distinct motional states. Furthermore, by dynamically controlling the measurement position, we achieve measurement-induced directional transport of a single atom without imparting additional momentum. Our results provide a direct experimental demonstration of the QZE in real space and establish a versatile framework for measurement-based control of atomic motion, paving the way for motional-state engineering in cold-atom platforms.}

The quantum Zeno effect (QZE)~\cite{misr77,PhysRevA.41.2295,PhysRevA.50.4582,PhysRevLett.109.150410,PhysRevLett.110.240403} refers to the inhibition of coherent evolution in a quantum system through frequent measurements~\cite{yan13,PhysRevLett.110.035302,PhysRevLett.112.070404}, effectively freezing the quantum state by repeatedly projecting it onto a given subspace~\cite{sch14,sign14,PhysRevLett.112.120406}. Originally proposed within the framework of ideal projective measurements, the concept was later generalized to scenarios involving weak measurements~\cite{PhysRevLett.60.1351,PhysRevLett.93.163602,PhysRevLett.94.220405,science.1226897} and continuous monitoring~\cite{PhysRevA.61.042107,mine19}, thereby establishing a profound connection between measurement, decoherence, and quantum state evolution~\cite{PhysRevLett.87.270405,PhysRevLett.93.130406}. The QZE was first predicted within the idealized framework of instantaneous orthogonal projections~\cite{misr77}. Later its theoretical description has been extended to a much broader scope, including continuous measurements, non-ideal weak monitoring, and non-Hermitian formulations, in all of which the QZE emerges naturally. Building on these developments, the von Neumann measurement model~\cite{veum55} explicitly incorporates the measuring apparatus as a quantum degree of freedom, describing the influence of frequent measurements on system dynamics through the interaction between the system and a probe. This approach provides a unified microscopic description of measurement processes and their backaction, and it naturally accounts for the complementary anti-Zeno effect~\cite{kofm00,PhysRevLett.87.040402,PhysRevLett.92.200403}, yielding a comprehensive theory that connects measurement frequency, measurement strength, and the intrinsic evolution of the system.

Experimentally, the QZE has been confirmed on a variety of physical platforms~\cite{PhysRevA.41.2295,yan13,bret15,PhysRevA.99.052101}. In trapped-ion systems, frequent resonant optical probing has been used to suppress coherent internal-state transitions, constituting the first direct observation of the QZE~\cite{PhysRevA.41.2295}. In superconducting qubit systems, continuous weak measurements via microwave cavities have enabled controllable switching between the QZE and anti-Zeno regimes~\cite{PhysRevLett.118.240401}. Furthermore, suppression of internal-state transitions has been observed in ultracold atomic gases~\cite{PhysRevLett.97.260402} and photonic systems~\cite{PhysRevLett.74.4763,PhysRevLett.83.4725}. These experiments, conducted in finite-dimensional Hilbert spaces, not only confirm the central predictions of QZE theory but also demonstrate how measurement-induced constraints can be employed to engineer effective Hamiltonians.

\begin{figure*}[ht]
    \centering
    \includegraphics[width=1.0\linewidth]{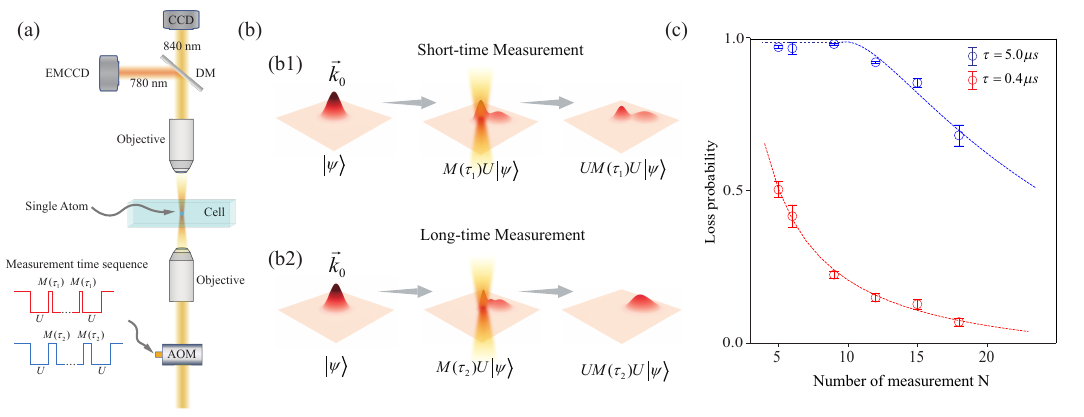}
    \caption{\textbf{Physical diagram and observation of the Quantum Zeno Effect.} (a) Experimental setup. An 840-nm laser, controlled by an acousto-optic modulator (AOM), is tightly focused by a high-NA objective into a vacuum cell to form an optical dipole trap for a single atom from a magneto-optical trap. Modulation of the AOM allows atom release or dipole pulses as measurements. Atomic fluorescence at 780-nm is collected by another objective and separated via a dichroic mirror. Two examples of modulation signals for short and long measurement pulses are shown on the left of the AOM. (b) Long- and short-time measurement. Top: A short pulse projects the motional state onto a trap eigenstate, leading to partial localization. Bottom: A long pulse projects the atom onto a trap eigenstate followed by in-trap dynamics. (c) Observation of the QZE. Multiple short or long measurement pulses are applied during a $ 45 ~ \mu \text{s} $ free evolution. For short pulses (red circles), the atom loss probability decreases against with measurement number, and the data are fitted by $ \text{P} = 2.99/\text{N} - 0.09$ (red dashed line). At a high measurement frequency, the loss probability approaches zero, indicating frozen evolutions. For long pulses (blue circles), the dynamics in dipole trap give rise to a clear difference compared to the short-pulse case. The decreasing scaling for measurement number N$>$10 is fitted by $ \text{P} = -136.67 (1/\text{N} - 0.1)^{2} + 0.99$. The error bars represent the standard deviation.}
    \label{fig:QZE_Diagram}
\end{figure*}

In contrast, experimental realizations of the QZE in continuous real space, where measurements directly probe the external motional degrees of freedom~\cite{klec06,kuba09,wils15,ross18}, remain relatively limited because of the stringent requirements on measurement frequency, strength, and spatial resolution. Nevertheless, some progress has been made in systems with external confinement. For example, in optical lattices, repeated high-resolution in-situ imaging of ultracold atoms has been used to suppress tunneling and diffusion within the lattice~\cite{PhysRevLett.115.140402}. However, for particles evolving in free space, where the free time evolution of the wave function directly competes with measurement backaction, systematic experimental investigations of the QZE remain largely unexplored. It is also worth emphasizing that measurements in experiments typically have a finite duration rather than being idealized instantaneous projections. In such cases, the system becomes effectively coupled to the detector, undergoing unavoidable evolution during the measurement process, which modifies the simple freezing effect expected from frequent projections. When the measurement duration is non-negligible, the influence on system at different time scales is still insufficiently understood and calls for further investigation.

In this work, we conduct a systematic experimental study on the influence of projective measurements on the free-space evolution of a single atom. A single atom, initially confined in an optical dipole trap, is released into the vacuum cell for free evolution, during which the activation of the dipole trap serves as a projective measurement. At the end of the evolution, in-situ imaging at the trap position is performed to determine whether the atom is lost. By repeating the experiment and statistically analyzing the loss probability, we obtain a direct characterization of the atomic dynamics in real space. We first measure the dependence of the atomic motion on the measurement frequency and, through numerical fitting, predict its behavior in the high-frequency measurement regime. By continuously varying the activation duration of the optical dipole trap, we systematically investigate the backaction of finite-duration measurements on the system’s evolution, thereby providing a clear physical picture of measurement backaction across different time scales. Moreover, we demonstrate that the cooperative influence of measurement frequency, strength, and spatial position enables precise regulation of the atomic motional states and can even induce directional atomic transport.

\subsection*{Physical model of measurement}

\begin{figure*}[ht]
    \centering
    \includegraphics[width=1.0\linewidth]{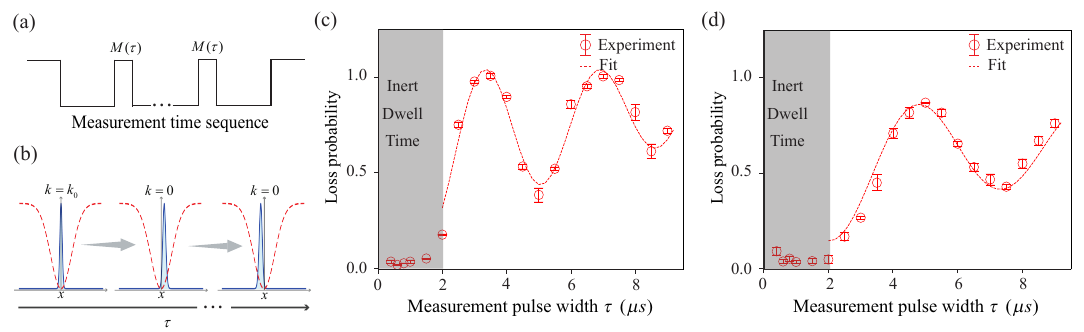}
    \caption{\textbf{Dynamics of a single atom under finite-time measurements.} (a) Measurement time sequence used in the experiment. The total free evolution time is $ 30 ~ \mu \text{s} $, with $ \text{N} = 15 $ inserted measurements. The pulse duration is varied to probe the motional dynamics in the trap. (b) Schematic of atomic dynamics in a one-dimensional harmonic trap. The atomic wavepacket oscillates periodically: starting at the trap center with wavevector $k=k_0$, its displacement from the center is maximal when $k=0$, and after half a period it reaches the opposite side. (c) Atom loss probability versus measurement pulse width. For pulse widths below $\sim 2.0 ~ \mu \text{s} $, the loss probability remains nearly zero, corresponding to the Inert-Dwell-Time regime. For longer pulses, the loss probability exhibits periodic oscillations, reflecting in-trap dynamics. The red dashed curve is a damped oscillatory fit $ \text{P} = 0.5 e^{-0.11 \tau }  \text{sin}(1.77\tau + 1.94) + ( 1 - 0.5 e^{-0.11 \tau } ) $. (d) Atom loss with reduced pulse intensity (half of (c)). The Inert-Dwell-Time regime remains $\sim 2.0 ~ \mu \text{s}$, while the oscillation period increases as the trap depth decreases, fitted by $ \text{P} = 0.415 e^{-0.09 \tau } \text{sin}(1.19\tau + 2.12 ) + 0.83 ( 1 - 0.5 e^{-0.11 \tau } ).$ The error bars represent the standard deviation.}
    \label{fig:Traped evolution}
\end{figure*}

Our physical model is based on the experimental procedure illustrated in Fig.~\ref{fig:QZE_Diagram}(a). A tightly focused optical tweezer is formed by an 840-nm laser beam, which is controlled by an acousto-optic modulator (AOM) and focused through a high-NA objective into a magneto-optical trap to capture a single cold atom as an optical dipole trap~\cite{schl01,PhysRevLett.89.023005}. The atom is then released into a vacuum cell for free evolution, during which the optical tweezer is intermittently switched on for a certain duration to serve as a projective measurement. Finally, in-situ imaging at the tweezer position is performed to determine whether the atom is lost. 

In this framework, the initial state of the atom is denoted as $\rho_0 = \ket{\phi_0} \bra{\phi_0}$, where $\ket{\phi_0}$ represents the atomic wavefunction immediately after release from the dipole trap, which follows a Gaussian distribution in real space. The subsequent motion inside the vacuum cell is governed by the free-particle Hamiltonian $H_{\mathrm{free}} = p^2/2m$ with the corresponding time-evolution operator $U_{\mathrm{free}}(t) = e^{-i H_{\mathrm{free}} t / \hbar}$. The action of the optical tweezer when switched on for a duration $\tau$ is described by 
\begin{equation}
\setlength\abovedisplayskip{4pt}
\setlength\belowdisplayskip{4pt}
    M(\tau) = U_{\mathrm{trap}} M_b,
\end{equation}
where $M_b=\ket{r} \bra{r}$ is the projection operator accounting for the measurement-induced collapse and $\ket{r}$ is the spatial window function corresponding to the localized state in the finite-depth potential of the tweezer. After the measurement, the density matrix updates according to:
\begin{equation}
\setlength\abovedisplayskip{4pt}
\setlength\belowdisplayskip{4pt}
    \rho \; \xrightarrow{\text{collapse}} \; \frac{M \rho M^\dagger}{\mathrm{Tr}[M \rho M^\dagger]}.
\end{equation}
Once projected, the atom again experiences the dipole trap, and its subsequent dynamics are governed by the trap Hamiltonian $H_{\mathrm{trap}} = \frac{p^2}{2m} - \frac{1}{2} m \Omega^2 r^2,$ with the corresponding propagator $U_{\mathrm{trap}}(t) = e^{-i H_{\mathrm{trap}} t / \hbar}$. Combining these elements, a single cycle of free evolution–measurement–trap evolution can be expressed as
\begin{equation}
\setlength\abovedisplayskip{4pt}
\setlength\belowdisplayskip{4pt}
    \rho_{k+1} = U_{\mathrm{trap}} \left( \frac{ M_b U_{\mathrm{free}} \rho_k U_{\mathrm{free}}^\dagger M_b^\dagger }{ \mathrm{Tr}[ M_b U_{\mathrm{free}} \rho_k U_{\mathrm{free}}^\dagger M_b^\dagger ] } \right) U_{\mathrm{trap}}^\dagger.
\end{equation}
After N such measurements, the atomic state becomes $\rho_\text{N}$, and the corresponding loss probability is
\begin{equation}
\setlength\abovedisplayskip{4pt}
\setlength\belowdisplayskip{4pt}
    \text{P}_{\mathrm{loss}} = \mathrm{Tr}\big[(1-M_b)\rho_{\text{N}}\big].
\end{equation}
This model not only provides an intuitive description of the measurement-induced wavefunction collapse, but also naturally incorporates the dynamical modifications arising from atomic motion inside the trap.  

\begin{figure*}[ht]
    \centering
    \includegraphics[width=1.0\linewidth]{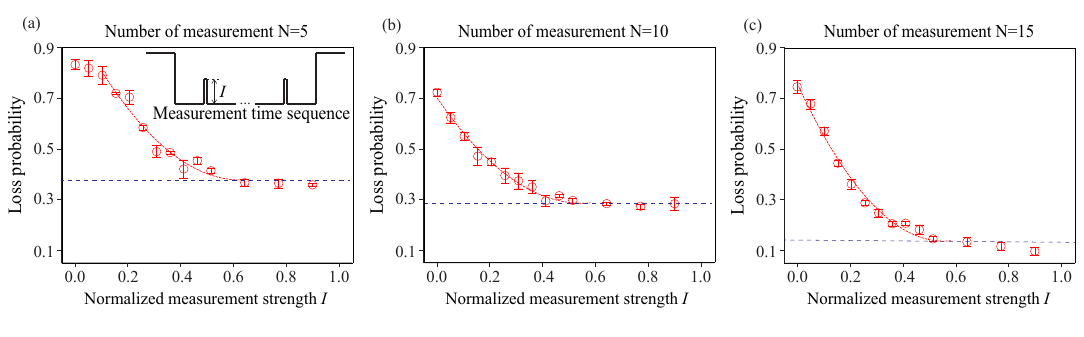}
    \caption{\textbf{Evolution under different measurement strengths.} During a $ 30 ~ \mu \text{s} $ free evolution, N equally spaced measurement pulses of $ 0.4 ~ \mu \text{s} $ duration are inserted. Panels (a)–(c) show the final atom loss probability versus measurement strength for $\text{N}=5,10,15$, respectively. The inset in (a) displays measurement time sequence used in the experiment, where the level $I$ means the measurement strength. For normalized strength below $ \sim 0.6 $, the loss probability decreases markedly with increasing strength, while above $ \sim 0.6 $ the decrease becomes much slower, thus defining the strong measurement regime. The decreasing scaling is fitted by a quadratic function $ \text{P} = 1.56 (I - 0.63 )^{2} + 0.37$($ \text{N} = 5 $), $ \text{P} = 1.40 (I -0.55 )^{2} + 0.28$($ \text{N} = 10 $), $ \text{P} = 2.05 (I - 0.55)^{2} + 0.14$($ \text{N} = 15 $), respectively. The error bars represent the standard deviation.}
    \label{fig:Strong Frequent measure}
\end{figure*}

\subsection*{Quantum Zeno effect of a single atom}
In the experiment, we fix the total free-space evolution time of the atom at $ \text{T} = 45 ~ \mu \text{s} $ and insert N measurements of duration $\tau$ at equal intervals during the evolution to systematically investigate the influence of finite-duration measurements on the atomic dynamics. We compare two typical cases: short time measurements ($\tau=0.4 ~ \mu\text{s}$) and long time measurements ($\tau=5.0 ~ \mu\text{s}$). In the short time case (Fig.~\ref{fig:QZE_Diagram}(b1)), after the wavefunction collapse, the atom is mainly localized within the region of the optical dipole trap, with only minor leakage along the initial momentum direction. Once the trap is switched off, the atom continues its directed motion and gradually expands into free space. In this regime, the measurement operator can be approximated as $M(\tau)\approx M_b$. 
Considering that the loss probability for a single evolution of duration $\text{T}  / \text{N}$ is $| \bra{r} H _{\text{free}} \ket{\phi_0} | ^{2} \times \text{T} ^{2} / ( \hbar ^{2} \text{N} ^{2} ) $, the cumulative loss probability after N measurements is proportional to $1/\text{N}$ [Details in Methods]. This behavior is illustrated by the red curve in Fig.~\ref{fig:QZE_Diagram}(c), the atomic loss probability rapidly approaches zero inversely with the number of measurements, indicating that frequent short measurements effectively freeze the evolution of a single atom, giving rise to the QZE.

In contrast, long-time measurements trigger pronounce intrap dynamics following wavefunction collapse (Fig.~\ref{fig:QZE_Diagram}(b2)), resulting in an increased wavefunction distribution outside the trap at the end of the measurement. As the atom can acquire additional momentum while confined, its subsequent free-space evolution strongly depends on the residence time inside the trap. This effect is exemplified by the blue curve in Fig.~\ref{fig:QZE_Diagram}(c), where for $\tau = 5.0 ~ \mu\text{s}$ the loss probability still decreases with increasing measurement frequency, yet remains higher than in the corresponding short-duration case. This indicates that the trap-induced dynamics $U_{\text{trap}}(\tau)$ dominates the measurement backaction on the atom. The influence of measurements on atomic motion across different timescales warrants further investigation.

\subsection*{Measurement-induced periodic dynamics}

To explore the influence of finite-duration measurements on atomic motion across different timescales, we continuously vary the measurement pulse width and quantitatively characterized the atomic dynamics in the optical tweezer potential by monitoring the loss rate. The experimental sequence is shown in Fig.~\ref{fig:Traped evolution}(a): $\text{N}=15$ measurement pulses of duration $\tau$ are inserted into a $30 ~ \mu \text{s}$ free-evolution window. Under the action of an effective projective measurement, the atomic wavefunction first collapses into the spatial extent of the dipole trap, with its wave packet centered at the potential minimum. Within the harmonic approximation~\cite{PhysRevA.47.R4567,PhysRevA.78.033425}, the atomic wavefunction subsequently undergoes periodic oscillations driven by the trap potential, while gradually losing momentum as its wave packet approaches the effective trap boundary until the wave vector reaches $k=0$, as schematically illustrated in Fig.~\ref{fig:Traped evolution}(b). Since the collapsed wave packet is centered at the harmonic origin (corresponding to zero phase) and the oscillation period is independent of the initial atomic velocity, the atom reaches its farthest displacement from the trap center (corresponding to the maximal loss probability) at the same time in each experimental run.

The experimental results corresponding to the above process are shown in Fig.~\ref{fig:Traped evolution}(c). When the measurement pulse duration is shorter than $\sim 2.0 ~ \mu \text{s}$, the atom exhibits negligible loss throughout the experiment. This characteristic timescale is independent of the trap parameters but decreases with longer free-evolution windows between trap switch-off events. At the critical value, the trap-induced spatial displacement of the wavefunction, combined with the subsequent free-space expansion, is just sufficient to drive the atom beyond the trapping region. By contrast, within this threshold the trap-induced displacement remains too small for the atom to escape. We refer to this timescale as the Inert Dwell Time. Once the pulse duration exceeds this threshold, the loss probability exhibits damped oscillations as a function of measurement duration, consistent with expectations from the harmonic approximation. The observed damping originates partly from the finite lifetime of atoms in the trap and partly from higher-order axial modes beyond the harmonic approximation, both of which effectively modulate the oscillation amplitude. A sinusoidal fit with an exponential envelope provides a good description of the experimental data. 

Furthermore, by reducing the measurement pulse intensity by half, we tune the effective trap frequency. As shown in Fig.~\ref{fig:Traped evolution}(d), the Inert Dwell Time ($\sim 2.0 ~ \mu \text{s}$) remains unchanged, while the oscillation frequency is significantly reduced compared with the high-intensity case. These results demonstrate that finite-duration position measurements of a single atom are governed by distinct physical mechanisms across different timescales. Short pulses act as near-projective probes, collapsing the atomic wavefunction into localized eigenstates, whereas long pulses induce coherent in-trap oscillations that lead to quantitative displacements of the wavefunction. Such timescale-dependent control indicates the potential of measurement as a powerful resource for preparing and manipulating motional quantum states~\cite{syas08,murc13,PhysRevX.5.041037,clar17}.

\subsection*{Crossover of measurement strength}

The strength of measurement plays a pivotal role in shaping the dynamical response of a quantum system. In general, sufficiently weak measurements exert only limited perturbations on the motional state, whereas strong measurements enforce wavefunction collapse through orthogonal projection. To investigate how dipole-trap measurements of varying strength affect the motion of a single atom, we tune the measurement strength by varying the pulse intensity and monitor the resulting motional dynamics during free evolution.

The measured atom-loss probabilities as a function of measurement strength are shown in Fig.~\ref{fig:Strong Frequent measure}(a–c). The inset of Fig.~\ref{fig:Strong Frequent measure}(a) illustrates the experimental time sequence: N measurement pulses of $0.4~\mu\text{s}$ duration are embedded within a $30~\mu\text{s}$ free-evolution window. At this timescale, the trap-induced unitary evolution $U_{\text{trap}}$ is negligible, allowing the pulses to be regarded as instantaneous measurements. Despite the different saturation values of loss probability under varying measurement frequencies, the dependence on measurement strength follows a universal trend. In the weak-measurement regime, each pulse perturbs the motional state, effectively reducing the outward propagation rate of the wavefunction and thus slightly lowering the loss probability compared to free evolution. This suppression follows a quadratic scaling with measurement strength until the threshold of about $0.4$ is reached. Beyond this threshold, further increasing the pulse strength no longer modifies the outcome, as the backaction acts as a strong projection that deterministically collapses the wavefunction into bound eigenstates of the dipole trap. Under frequent strong measurements [Fig.~\ref{fig:Strong Frequent measure}(c)], the loss probability approaches zero, corresponding to a freezing of motional dynamics induced by the quantum Zeno effect. These observations reveal a clear crossover from weak-measurement-induced motional perturbations to strong-measurement-induced collapse, and demonstrate that joint control over measurement strength and frequency provides a versatile means to engineer atomic motional quantum states~\cite{katz06,PhysRevLett.104.080503}.

\subsection*{Manipulation of motional state}

\begin{figure}[h]
    \centering
    \includegraphics[width=1.0\linewidth]{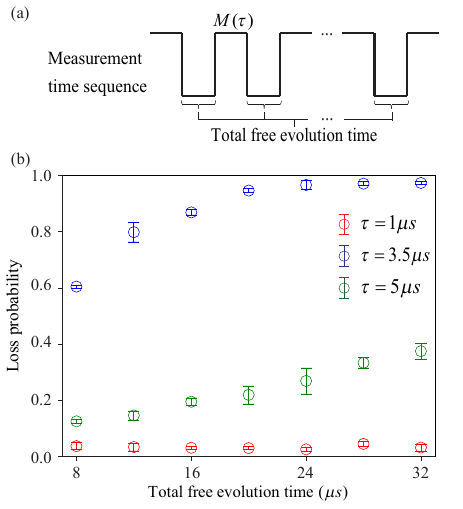}
    \caption{\textbf{Free evolution interrupted by repeated measurements.} (a) Experimental sequence. During free evolution, measurement pulses are inserted every $2 ~ \mu \text{s}$, with pulse widths set to $ 1 ~ \mu \text{s} $, $ 3.5 ~ \mu \text{s} $, and $5 ~ \mu \text{s} $, respectively. (b) Atom loss probability as a function of the total free evolution time for different pulse durations. For $ 1 ~ \mu \text{s}$ pulses (red open circles), the loss probability remains negligible and independent of the evolution time. For $ 3.5 ~ \mu \text{s} $ pulses (blue open circles), the loss probability increases rapidly with evolution time and approaches nearly unity beyond $ 24 ~ \mu \text{s} $. In contrast, for $ 5 ~ \mu \text{s} $ pulses (green open circles), the loss probability rises much more slowly and stays below $0.4$ even at $32 ~ \mu \text{s} $, indicating a substantial suppression of atom loss compared with the $ 3.5 ~ \mu \text{s} $ case. The error bars represent the standard deviation.}
    \label{fig:Motional State Engineering}
\end{figure}

\begin{figure*}[ht]
    \centering
    \includegraphics[width=1.0\linewidth]{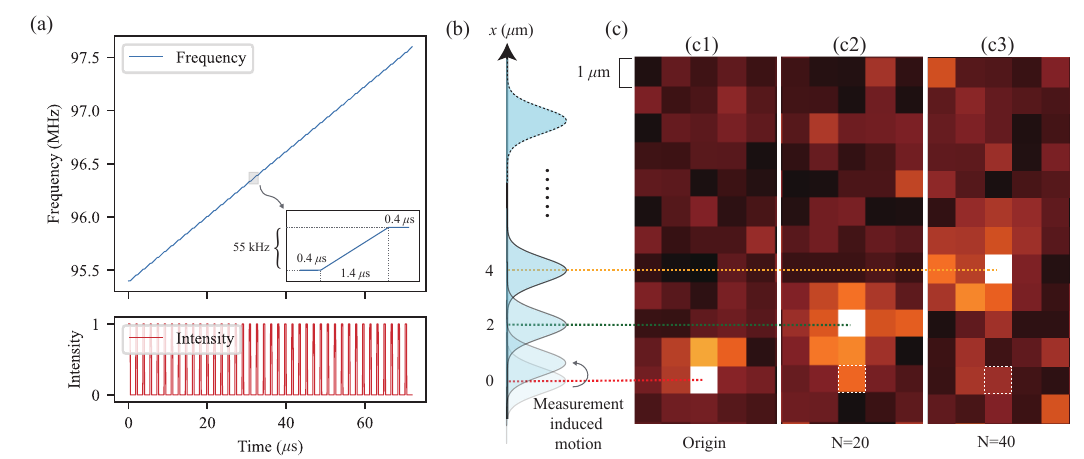}
    \caption{\textbf{Controlled atomic motion via frequent measurements.} (a) Measurement time sequence. The upper panel shows the frequency control sequence of the acousto-optic deflector (AOD) driving signal, with the frequency increasing stepwise from $95.4 ~ \text{MHz}$. The lower panel shows the intensity control sequence of the AOD driving signal, corresponding to the timing of the applied measurements. The measurement pulse width is $0.4 ~ \mu \text{s}$, with a $1.4 ~ \mu \text{s}$ interval between measurements. The inset in the upper panel shows that the frequency variation of $ 55 ~ \text{kHz} $ for every step occurs within the $1.4 ~ \mu \text{s}$ interval when the intensity is zero. (b) Schematic of atomic wave packet motion. Each measurement pulse induces a slight shift in the center position relative to the previous one, resulting in a corresponding displacement of the atomic wave packet’s center after projection. After multiple such measurements, the center of the atomic wave packet undergoes a noticeable shift. (c) Experimental results. (c1) Atomic position imaging before any measurements are applied. (c2) Atomic position imaging after $ \text{N} = 20 $ measurements, showing a displacement of approximately $ 2 ~ \mu \text{m} $ from the initial position. (c3) Atomic position imaging after $ \text{N} = 40 $ measurements, showing a displacement of approximately $ 4 ~ \mu \text{m} $ from the initial position. Both displacements are measured relative to the position before measurements were applied.}
    \label{fig:Movement}
\end{figure*}

Building on the above results, we combine control over the duration, strength, and frequency of the measurement pulses with adjustment of the total free-evolution window to demonstrate directional control of the atomic motional state. The experimental sequence is shown in Fig.~\ref{fig:Motional State Engineering}(a): during each free-evolution interval, measurement pulses are applied every $2~\mu\text{s}$. By extending the total free-evolution time while keeping the pulse spacing fixed, we monitor the motion of atom under different measurement conditions via the loss probability.

When suppression of atomic motion is desired, frequent short-pulse measurements strongly localize the wave packet near the trap center, enabling the atom to remain in a low-loss motional state even over long free-evolution durations (Fig.~\ref{fig:Motional State Engineering}(b), red points). In contrast, when the objective is to facilitate the atom’s escape from the trap, strong measurements with pulse durations matched to integer multiples of the oscillation period are applied. In this case, the atomic wave packet undergoes significant displacement within the potential, gaining additional energy through repeated switching of the dipole trap. As a result, the atom acquires larger momentum and can rapidly leave the bound state within a short free-evolution interval (Fig.~\ref{fig:Motional State Engineering}(b), blue curve). Interestingly, when strong measurements are synchronized with a phase of $\pi/2$ relative to the unbound motional state, the oscillatory nature of the trap partially restores spatial localization. Experimentally, this manifests as a revival of low-loss behavior at specific evolution times, as shown by the green points in Fig.~\ref{fig:Motional State Engineering}(b). This condition corresponds to a control duration extended by half a period relative to $\tau = 3.5~\mu\text{s}$. These results clearly establish measurement as a flexible resource for tailoring the motional state~\cite{wise09,Jaco14}, with the potential to induce directed atomic motion when appropriately designed.

\subsection*{Directional atomic motion}
Building upon in-situ measurements of the atom, we further introduce a series of projective measurements with gradually shifted spatial positions. Experimentally, this scheme can be realized by replacing the AOM in the optical path with an acousto-optic deflector (AOD). By jointly controlling the frequency and intensity of the radio-frequency (RF) signal applied to the AOD, each measurement is displaced by a fixed offset relative to the previous one. The control sequence is illustrated in Fig.~\ref{fig:Movement}(a), where the inset depicts a single step of the process: after applying a $0.4 ~ \mu \text{s}$ projective strong measurement, the dipole trap is switched off (corresponding to zero RF intensity), followed by $1.4 ~ \mu \text{s}$ of free-space evolution. During this free evolution window, a frequency shift is applied to the RF drive. At the end of this interval, a signal with the same intensity but a frequency $55~\text{kHz}$ higher than the previous one is applied to the AOD. In our experimental system, this control sequence results in each measurement trap being displaced by approximately $0.1 ~ \mu \text{m}$ relative to the preceding one. Repeating this process $\text{N}$ times, the measurement backaction on the atom can be expressed as $M=(\prod _{j=0}^{N} M_j)^\dagger$, where $M_i=\ket{r+j\Delta r}\bra{r+j\Delta r}$ denotes the measurement at a position shifted by $j \Delta r$ relative to the origin.

Under such a sequence, the role of frequent measurements is no longer to freeze the atom at a fixed position. Instead, the atomic wavefunction is repeatedly projected in a directional manner, with the wave packet center being transferred step by step to the position of the dipole trap, i.e., measurement-induced directed motion of the atom, as illustrated in Fig.~\ref{fig:Movement}(b). Experimentally, we first perform in-situ imaging of the atom (Fig.~\ref{fig:Movement}(c1)), followed by global imaging for measurement numbers $\text{N} = 20$ (Fig.~\ref{fig:Movement}(c2)) and $\text{N} = 40$ (Fig.~\ref{fig:Movement}(c3)). The results clearly show the atom bound in the dipole trap displaced by approximately $2 ~ \mu \text{m}$ and $4 ~ \mu \text{m}$ from the initial position, respectively. Importantly, this displacement is not caused by optical tweezer dragging, but rather by the effect of high-frequency position measurements. In this process, the effective atomic drift velocity is about $0.7~\text{m/s}$, which surpasses the maximum dragging velocity of $0.05~\text{m/s}$ in the same experimental system. This approach not only introduces a novel paradigm for controlling atomic displacement~\cite{Kim2016}, but, in contrast to dragging-based techniques, it avoids imparting extraneous momentum to a single atom. Consequently, the atom remains in a low-energy motional state after displacement, underscoring the potential of this method as a powerful tool for coherent qubit manipulation.

\subsection*{Discussion and conclusion}
In this work, we employ short openings of an optical dipole trap as finite-duration projective measurements to systematically investigate the interplay between free evolution and measurement for a single cold atom in free space. By comparing the effects of frequent measurements with different durations, we directly observe the QZE in real space and explicitly demonstrate the evolution differences between finite-duration measurements and ideal instantaneous projections arising from measurement backaction. Furthermore, through data fitting, we predict the emergence of frozen dynamics under the regime of ultrafrequent measurements. We also discover that under measurement coupling, the atomic dynamics exhibit a distinctive structure: an initial nearly frozen stage followed by periodic motion. While the oscillation period is controlled by the dipole trap intensity, the duration of the frozen stage remains insensitive to trap strength and can thus be approximated as an instantaneous projection. This behavior provides a clear physical picture of finite-duration measurements, highlighting that real measurements are not a singular projection event but rather a process composed of dynamical stages on different time scales.

Building on this understanding, we further systematically examine the combined influence of measurement frequency and strength on the free evolution of a single atom. Our experimental results demonstrate that measurements can not only suppress or delay the spatial spreading of the quantum state, but also deterministically prepare distinct motional states through appropriate tuning of measurement parameters. Beyond in-situ measurements, by controlling the spatial position of the applied measurements, we are able to directly induce directional motion of the atomic wavefunction, thereby realizing atomic transport without imparting additional momentum. Overall, our work provides a clear experimental demonstration of the QZE in real space for a freely evolving particle, elucidates the role of measurement in realistic experimental settings, and establishes a foundation for the broader exploration of measurement-induced dynamics~\cite{PhysRevB.100.134306,PhysRevX.11.011030,noel22}. These results open a feasible pathway toward motional state engineering in cold-atom platforms~\cite{RevModPhys.80.885,kauf15,jock15}.

\subsection*{Methods}

\noindent
\textbf{Quantum Zeno scaling of the atom loss probability.} We consider the free evolution of a single atom over a total duration $ \text{T} $, during which $ \text{N} $ projective measurements are inserted at equal intervals of $ \text{T/N} $. The atom is initially prepared in the state $ \ket{\phi_0} $ immediately after release from the trap. After a free evolution of duration $ \text{T/N} $, the state evolves to $ U_{\mathrm{free}}( \text{T/N} )\ket{\phi_0} = e^{-i H_{\mathrm{free}} \text{T} / ( \hbar \text{N} )} \ket{\phi_0} $. A measurement operator $M_b= \ket{r} \bra{r}$ is then applied, with $ \ket{r} = \ket{\phi_0} $ for the case that the trap characteristics remain unchanged. The post-measurement state thus becomes $ \bra{r} e^{-i H_{\mathrm{free}} \text{T} / \hbar \text{N} } \ket{\phi_0}  \ket{r} $, and the atom loss probability is $ \text{P}_{1} = 1 - | \bra{r} e^{-i H_{\mathrm{free}} \text{T} / \hbar \text{N} } \ket{\phi_0} | ^ {2} $. 

For sufficiently small $ \text{T/N} $, the propagator can be expanded to first order, yielding the approximation $ \bra{r} e^{-i H_{\mathrm{free}} \text{T} / \hbar \text{N} } \ket{\phi_0} \approx 1 - i \bra{r} H_{\mathrm{free}} \ket{\phi_0} \text{T} / ( \hbar \text{N} ) $. This leads to the expression of the single-measurement loss probability $ \text{P}_{1} = |\bra{r} H_{\mathrm{free}} \ket{\phi_0} | ^ {2} \text{T} ^{2} / (\hbar ^{2} \text{N} ^{2} ) $. Extending this to $\text{N}$ repeated measurements gives the final loss probability $ \text{P}_{ \text{N} } = 1 - [ 1 - |\bra{r} H_{\mathrm{free}} \ket{\phi_0} | ^ {2} \text{T} ^{2} / (\hbar ^{2} \text{N} ^{2} ) ] ^{ \text{N} } $. Taking the limit $ \text{N} \to \infty $, we obtain the first order approximation $ \text{P}_{ \text{N} } = |\bra{r} H_{\mathrm{free}} \ket{\phi_0} | ^ {2} \text{T} ^{2} / (\hbar ^{2} \text{N} ) $, which predicts that the atom loss probability decreases linearly with increasing measurement number.\\

\noindent
\textbf{Magneto-optical trap loading.} To stochastically load a single $^{87}$Rb atom into the dipole trap, a magneto-optical trap (MOT) is first established at the trap site. The MOT employs three intersecting retroreflected beams containing both cooling and repump light, together with an anti-Helmholtz magnetic field. One pair of beams is oriented perpendicular to the optical table, while the other two pairs are confined to lie within the plane of the table. Due to the short working distance ($\sim 15 ~ \text{mm}$) of the high–numerical aperture objective, the in-plane beams cannot intersect at the conventional $ 90 ^{\circ}$, but instead cross at an angle of $\sim 120 ^{\circ}$. Each cooling beam has a diameter of $\sim 20 ~ \text{mm}$ and a power of $\sim 6 ~ \text{mW} $, while each repump beam carries $\sim 1 ~ \text{mW} $. The magnetic field gradient is set to $\sim 14 ~ \text{ G/cm} $. The cooling light is red-detuned by $ 15 ~ \text{MHz} $ from the $ \ket{5S_{1/2}, F = 2} \to \ket{5P_{3/2}, F = 3} $ transition, and the repump light is resonant with the $ \ket{5S_{1/2}, F = 1} \to \ket{5P_{3/2}, F = 2} $ transition.

Rubidium atoms are supplied from a thermally heated dispenser, with atomic flux regulated by a tunable valve. For single-atom trapping, the valve is adjusted to its minimum opening, thereby reducing the MOT loading rate and suppressing background collisions that would otherwise limit the trap lifetime.\\

\begin{figure}[tb]
    \centering
    \includegraphics[width=0.95\linewidth]{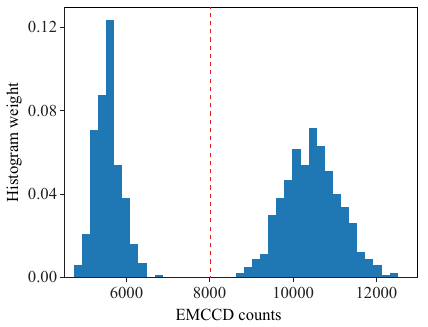}
    \caption{\textbf{Histogram of single-atom fluorescence counts.} The histogram shows the EMCCD count distribution from $1,000$ repeated fluorescence imaging experiments. The histogram exhibits a clear bimodal structure, with peaks centered at approximately $5500$ and $10500$ counts, corresponding to empty traps and single-atom occupancy, respectively. The red dashed line denotes the threshold used to discriminate successful single-atom loading.}
    \label{fig:Signal_statistic}
\end{figure}

\noindent
\textbf{Single-atom optical tweezer.} A single-atom optical trap is realized by tightly focusing an 840-nm far-off-resonant Gaussian beam using a custom high–numerical aperture ($\text{N.A.} = 0.5$) objective, yielding a beam waist of $ 1.1 ~ \mu \text{m} $. At this scale, collisional blockade~\cite{PhysRevLett.89.023005} ensures that the trap hosts at most one atom. The trapping beam power is set to $ 8 ~ \text{mW} $, corresponding to a trap depth of $ 2.3 ~ \text{mK} $, with a single-atom lifetime of several seconds under a vacuum level below $1 \times 10^{-8} ~ \text{Pa}$, limited primarily by off-resonant photon scattering and residual background gas collisions. The trapping beam is being intensity-controlled and rapidly switched using an acousto-optic modulator (AA MT80-B30A1-IR), which is driven by an arbitrary waveform generator to flexibly program the pulse sequences and intensities. \\

\noindent
\textbf{Fluorescence detection.} Trap occupancy is detected by fluorescence imaging. Two probe beams, one red-detuned by $ 30 ~ \text{MHz} $ from the cooling transition and another resonant with the repump transition, are applied perpendicular to the dipole-trap axis to minimize background scattering. Each probe beam has a diameter of $ \sim 1 ~ \text{mm} $; the cooling probe carries $\sim 0.4 ~ \text{mW}$ of power, while the repump probe carries $\sim 0.1 ~ \text{mW}$. Atomic fluorescence is collected by a second objective, spectrally separated from the 840-nm light by a dichroic mirror, and recorded on an electron-multiplying CCD (EMCCD, ANDOR iXon Ultra 897) with EM Gain = 100. 

To further suppress background contributions, the dipole-trap beam is spectrally filtered with three 840-nm bandpass filters before the high-NA objective, while three additional 780-nm bandpass filters are placed in front of the EMCCD to block stray light at other wavelengths. This filtering scheme is crucial for achieving a high signal-to-noise ratio, enabling unambiguous discrimination of single-atom fluorescence against residual trap light and scattered background.\\

\noindent
\textbf{Experimental time sequence.} The sequence for preparing a single atom proceeds as follows: the MOT and dipole-trap beams are switched on simultaneously. After $ 140 ~ \text{ms} $, the MOT beams and magnetic field are turned off, allowing a single atom to be loaded in the dipole trap during the ensemble decay. Following a 25-ms wait, the probe beams and EMCCD are activated for 20 ms to record the atomic fluorescence. The 25-ms delay ensures that the residual cold atomic cloud has sufficiently dispersed, thereby preventing background atoms from contributing to the subsequent fluorescence signal used to determine single-atom loading.

Once a single atom is prepared in the trap, the experimental measurement sequence is applied. After the completion of the measurement protocol, a second fluorescence image is acquired $ 40 ~ \text{ms} $ after the initial detection. The interval between the two images is chosen to accommodate the readout time of the EMCCD, while still allowing reliable detection of the atom at the end of the experimental cycle. The first fluorescence image serves to identify successful single-atom loading, whereas the second image provides the readout of the experimental outcome.\\

\noindent
\textbf{Single-atom discrimination.} 
The fluorescence signal is used to discriminate single-atom occupying events. To balance signal strength with acquisition time, the fluorescence imaging duration is set to $ 20 ~ \text{ms} $, which provides sufficient photon counts for reliable atom detection without introducing excessive delay to the experimental cycle. A histogram of photon counts from $1000$ experimental realizations displays a clear bimodal distribution, corresponding to zero or one atom in the trap (Fig.~\ref{fig:Signal_statistic}). The threshold between the two peaks defines the single-atom loading criterion. The statistical analysis yields a stochastic loading probability of $ 0.5 \sim 0.6 $. The absence of any higher-photon-count peak confirms the collisional blockade limit, ensuring that at most one atom is present in the trap.\\

\noindent
\textbf{Adjustable measurement pulse position control.} To realize measurement pulses with adjustable center positions, the dipole trap light passes through an acousto-optic deflector (AOD, AA DTSXY-400-800) before entering the high-numerical-aperture objective. By tuning the frequency and intensity of the AOD’s driving signal, we implement the required sequence for repeated measurements in the experiment. In this setup, the frequency difference between the driving signals of consecutive measurement pulses is $55 ~ \text{kHz}$, corresponding to a center position shift of approximately $0.1 ~ \mu \text{m}$. The width of each measurement pulse is set to $0.4 ~ \mu \text{s}$, and the free evolution time between consecutive measurements is $1.4 ~ \mu \text{s}$. During this free evolution time, the FPGA gradually adjusts the driving signal frequency of the measurement pulses without altering the intensity, ensuring that the measurement pulse intensity remains at zero and does not interfere with the atom's free evolution. 

Fig.~\ref{fig:Movement}(c) shows the experimental result of single-shot atomic fluorescence imaging signals captured by the EMCCD. The pixel size of the EMCCD is $16 ~ \mu \text{m}$, and considering the magnification of the imaging system, the actual distance in the plane where the atom stays corresponds to approximately $1 ~ \mu \text{m}$.

\section*{Acknowledgements}
We thank Prof. Wei Zhang for helpful discussions of single atoms. We acknowledge funding from the National Key R and D Program of China (Grant No. 2022YFA1404002), the National Natural Science Foundation of China (Grant Nos. T2495253, 62435018). 

\section*{Author contributions statement}
Z.Y.Z. and H.C.C. conducted the physical experiments and wrote the paper. All authors contributed to discussions regarding the results and the analysis contained in the manuscript. D.-S.D. conceived the idea and supported this project.

\section*{COMPETING INTERESTS}
The authors declare no competing interests.

\nocite{*}

\bibliography{citation}% Produces the bibliography via BibTeX.

\end{document}